**Ab-initio study of the bandgap engineering of $Al_{1-x}Ga_xN$ for optoelectronic applications**

*B. Amin[1], Iftikhar Ahmad[1,†], M. Maqbool[2], S. Goumri-Said[3], R. Ahmad[4]*

1. *Department of Physics, Hazara University, Mansehra, Pakistan*
2. *Department of Physics and Astronomy, Ball State University, Muncie, IN 47306, USA*
3. *KAUST, Saudi Arabia*
4. *Department of Chemistry, Hazara University, Mansehra, Pakistan*

## ABSTRACT

A theoretical study of $Al_{1-x}Ga_xN$, based on full-potential linearized augmented plane wave method, is used to investigate the variations in the bandgap, optical properties and non-linear behavior of the compound with the variation of Ga concentration. It is found that the bandgap decreases with the increase of Ga in $Al_{1-x}Ga_xN$. A maximum value of 5.5 eV is determined for the bandgap of pure AlN which reaches to minimum value of 3.0 eV when Al is completely replaced by Ga. The static index of refraction and dielectric constant decreases with the increase in bandgap of the material, assigning a high index of refraction to pure GaN when compared to pure AlN. The refractive index drops below 1 for photon energies larger than 14 eV results group velocity of the incident radiation higher than the vacuum velocity of light. This astonishing result shows that at higher energies the optical properties of the material shifts from linear to non-linear. Furthermore, frequency dependent reflectivity and absorption coefficients show that peak value of the absorption coefficient and reflectivity shifts towards lower energy in the UV spectrum with the increase in Ga concentration. This comprehensive theoretical study of the optoelectronic properties of the alloys is presented for the first time which predicts that the material can be effectively used in the optical devices working in the visible and UV spectrum.

†. Corresponding author email: ahma5532@gmail.com

# I. INTRODUCTION

Group-III nitride family and their alloys are suitable materials for optoelectronic applications in short wave length, high temperature, high frequency and high power electronic devices [1, 2]. Rare-earth doped AlN and GaN are effectively used in photonic devices, electroluminescent devices (ELDs), diode lasers, medicine and health physics [3-9]. Accurate knowledge of the bandgap and optical properties of these compounds is very important for the design and analysis of various optoelectronic and photonic devices. It is important to understand fundamental optical properties over a wide range of wavelengths. In reality, optical properties reflect the density of states (DOS) of a compound and their analysis is one of the most effective tools to understand the electronic structure of the material [10, 11].

$Al_{1-x}Ga_xN$ and $Al_{1-x}In_xN$ are well suited materials as photo-detectors in the UV spectrum [12]. Many experimental and theoretical studies have been performed on the wurtzite structure of these ternary alloys [13-15], but little work has been reported on the zinc-blend phase of these materials [16], even after the successful growth reported by Okumura et al. [16] and Martinez et al. [17]. The investigation of zinc-blend structure of these materials is more important than wurtzite structure due to its large optical gain and lower threshold current density [18].

Lee and Wang [19] studied the electronic structure of zinc-blend $Al_{1-x}Ga_xN$ by local density approximation (LDA) and screened-exchange local density approximation (sX-LDA). While Kanoun et al. [20] reported the structural parameters and bandgap bowing in zinc-blend phase of $Al_{1-x}Ga_xN$ and $Al_{1-x}In_xN$ using density functional theory with local density approximation (LDA). It has been confirmed that generalized gradient approximation (GGA) is far more superior to LDA in the band structure calculation [21, 22], that's why GGA is used in the present work. Though bandgap and electronic structure of $Al_{1-x}Ga_xN$ has already been studied in Ref. [19] and Ref. [20] but no theoretical work is reported on the optoelectronic properties of the compound. Optical properties of $Al_{1-x}Ga_xN$ and their dependence on the variation of Ga concentration are calculated.

## II. THEORY AND CALCULATION

The dielectric function $\varepsilon(\omega)$ of the electron gas, with its strong dependence on frequency has significant effect on the physical properties of solids. It describes the collective excitations of the Fermi Sea, such as the volume and surface plasmons. The dielectric function depends on the electronic band structure of a crystal and its investigation by optical spectroscopy is a powerful tool in the determination of the overall band behavior of a crystal. It can be divided into two parts; real and imaginary parts:

$$\varepsilon(\omega) = \varepsilon_1(\omega) + i\varepsilon_2(\omega) \quad (1)$$

The imaginary part of the complex dielectric function $\varepsilon_2(\omega)$ in cubic symmetry compounds can be calculated by the following relation [23, 24]:

$$\varepsilon_2(\omega) = \frac{8}{2\pi\omega^2} \sum_{nn'} \int_{BZ} |P_{nn'}(k)|^2 \frac{dS_k}{\nabla \omega_{nn'}(k)} \quad (2)$$

$\varepsilon_2(\omega)$ is strongly dependent on the joint density of states, $\omega_{nn'}$, and momentum matrix element, $P_{nn'}$. While Kramers-Kronig relation is used to calculate the real part of the dielectric function $\varepsilon_1(\omega)$ from $\varepsilon_2(\omega)$ [25]:

$$\varepsilon_1(\omega) = 1 + \frac{2}{\pi} P \int_0^\infty \frac{\omega' \varepsilon_2(\omega')}{\omega'^2 - \omega^2} d\omega' \quad (3)$$

The values of the real and imaginary parts of the frequency dependent dielectric function provides basis for the calculation of the refractive indices $\tilde{n}(\omega)$. The refractive indices of ternary alloy are important in the designing of the optical parameter [26]. The complex refractive index of a compound is:

$$\tilde{n}(\omega) = n(\omega) + ik(\omega) = \varepsilon^{1/2} = (\varepsilon_1 + i\varepsilon_2)^{1/2} \tag{4}$$

Where $n(\omega)$ represent the real part of the refractive index and $k(\omega)$ represent the imaginary part or the extinction coefficient [27]. The real part of the refractive index $n(\omega)$ and the extinction coefficient $k(\omega)$ can be calculated from $\varepsilon_1(\omega)$ and $\varepsilon_2(\omega)$ using the following equations:

$$n(\omega) = \frac{1}{\sqrt{2}}\left[\left\{\varepsilon_1(\omega)^2 + \varepsilon_2(\omega)^2\right\}^{1/2} + \varepsilon_1(\omega)\right]^{1/2} \tag{5}$$

$$k(\omega) = \frac{1}{\sqrt{2}}\left[\left\{\varepsilon_1(\omega)^2 + \varepsilon_2(\omega)^2\right\}^{1/2} - \varepsilon_1(\omega)\right]^{1/2} \tag{6}$$

Using the above optical parameters $n(\omega)$ and $k(\omega)$; the frequency dependent reflectivity, $R(\omega)$ can be calculated by the following relation:

$$R(\omega) = \left|\frac{\tilde{n} - 1}{\tilde{n} + 1}\right| = \frac{(n-1)^2 + k^2}{(n+1)^2 + k^2} \tag{7}$$

Similarly the absorption coefficient is given by Beer's law [27]:

$$\alpha = \frac{2k\omega}{c} = \frac{4\pi k}{c} \tag{8}$$

The absorption coefficient can also be calculated from the dielectric function [27]:

$$\alpha(\omega) = 2\omega k(\omega) = \sqrt{2}\omega\left[\left\{\varepsilon_1(\omega)^2 + \varepsilon_2(\omega)^2\right\}^{1/2} - \varepsilon_1(\omega)\right]^{1/2} \tag{9}$$

In this article, all density functional calculations are performed with the full-potential linearized augmented plane wave method, using Wu-Cohen potential in the generalized gradient approximation embedded in the wien2k software [28]. The details of calculation are reported in our previous work [5, 6, 24, 24, 29]. In the present work we used $R_{MT} * K_{max} = 8.00$ value for the plane wave cut-off in the interstitial region. For the calculation of the optical properties of $Al_{1-x}Ga_xN$; a mesh of 3500 k-points is taken for the Brillouin zone integrations in the corresponding irreducible wedge.

## III. RESULTS AND DISCUSSION

Electronic bandgaps, optical parameters such as dielectric constants and refractive indices for $Al_{1-x}Ga_xN$ are calculated. Fig.1 shows that the calculated bandgap varies linearly as a function of Ga concentration. Bandgap is of direct nature for the full compositional range of Ga in $Al_{1-x}Ga_xN$ ($0 \leq x \leq 1$). For x = 0 with no Ga concentration, the bandgap of pure AlN is determined to be 5.50 eV. By increasing the fractional concentration x of Ga, the bandgap of the materials decreases. This decrease saturates at the maximum value of x and gives a bandgap of 3.00 eV for pure GaN. Moreover, our calculated results of the bandgaps for $Al_{1-x}Ga_xN$ ($0 \leq x \leq 1$) is shown in Fig.1, are closer to the experimental results of AlN and GaN [30, 31]. It is further noted that our theoretical results are much improved than reference [20]. The reason for our better results is using Wu-Cohen potential in the GGA scheme [32] which is more effective in the bandgap calculation than the reported results of LDA [21, 22].

The variation in the bandgap with the constituents' concentration occurs due to the difference in the density of states [6]. It has already been reported that the density of states affects the band structure of a compound and hence its bandgap [33-35]. Furthermore, difference in the densities of states of AlN and GaN are responsible for making AlN as a wider bandgap material than GaN. When some of the Al atoms are replaced by Ga atoms in the AlN crystal, then the mixing of Al and Ga in the compound is responsible for the change in the density of states. At a higher Al concentration the distribution of the density of states of Al is dominant, making the bandgap of the resulting material closer to the bandgap of pure AlN. In a similar way at higher Ga concentration, the bandgap of $Al_{1-x}Ga_xN$ is closer to pure GaN due to the dominant effect of the distribution of the density of states of GaN.

The variation in the bandgap of $Al_{1-x}Ga_xN$ provides a promising result of the bandgap engineering for optoelectronic devices. It covers a full spectrum of wavelengths from visible to UV, because the bandgap of pure AlN is transparent to visible light spectrum and part of the ultraviolet light while pure GaN is transparent to visible light. Depending upon the need and requirement of a particular application, any desired bandgap in this material can be achieved within a limit of 3.0 eV to 5.50 eV. Thus by varying the Ga concentration in $Al_{1-x}Ga_xN$, optoelectronic devices in various regions of the spectrum can be formed. It can also be used to construct quantum wells of a desired width and depth [6, 36, 37].

The calculated real and imaginary parts of the dielectric function of $Al_{1-x}Ga_xN$ at x = 0, 0.25, 0.50, 0.75 and 1.0 for the energy range 0 to 25 eV are shown in Fig.2. It is clear from the figure that for x = 0, 0.25, 0.50, 0.75 and 1.0 the critical points in the imaginary parts of the dielectric function occurs at about 5.43 eV, 4.70 eV, 4.19 eV, 3.53 eV and 2.86 eV. These points are closely related to the direct band gaps ($E_g^{\Gamma-\Gamma}$ ); 5.50 eV, 4.78 eV, 4.30 eV, 3.60 eV and 3.0 eV of $Al_{1-x}Ga_xN$ at x = 0, 0.25, 0.50, 0.75 and 1.0. It is clear from the figure that AlN have strong absorption region 7.01 eV to 15.35 eV. This region consists of different peaks. The width of the absorption region decreases by changing the concentration of Ga from 0 to 100 % and the sharpness and height of the peaks increases. The variations in the linear optical absorption region from one compound to another can be related to the bandgap of the materials. The materials with a bandgap lesser than 3.1 eV work well in the visible light devices applications while those with bandgap larger than 3.1 eV can also be used in ultraviolet (UV) applications [6-9, 5]. These prominent variations in the optical absorption region with bandgap of $Al_{1-x}Ga_xN$ (3.0 eV to 5.50 eV) confirms its suitability for device applications in the major parts of the spectrum; visible to UV.

Fig. 2 also show the calculated real parts of the complex dielectric function $\varepsilon_1(\omega)$ for $Al_{1-x}Ga_xN$. It is clear from the figure that the static dielectric constant $\varepsilon_1(0)$, the low energy limit of $\varepsilon_1(\omega)$, is strongly dependent on the bandgap of a compound. The calculated values of $\varepsilon_1(0)$ Vs bandgap is plotted in Fig.3. The plot shows inverse relation of $\varepsilon_1(0)$ with the bandgap. The inverse relation of $\varepsilon_1(0)$ with the bandgap can be explained by Penn Model [38].

$$\varepsilon(0) \approx 1 + (\hbar\omega_p / E_g)^2 \tag{10}$$

The above relation can be used to calculate $E_g$ using the value of $\varepsilon_1(0)$ and plasma energy $\hbar\omega_p$. It is evident from Fig.2 that the first hump for AlN appears at 7.16 eV. The hump is shifted toward higher energies by increasing the concentration of Ga.

For all the investigated compounds of $Al_{1-x}Ga_xN$ at x = 0, 0.25, 0.50, 0.75 and 1.0; $\varepsilon_1(\omega)$ becomes zero at certain energy and then decreases to minimum values, with negative numbers, about 14.98 eV, 14.35 eV, 14.21 eV, 13.58 eV and 13.45 eV. The negative values of $\varepsilon_1(\omega)$ show that in this energy region the incident electromagnetic waves are totally reflected from the medium, hence the material exhibit metallic nature. The negative values of $\varepsilon_1(\omega)$ correspond to the local maxima of the reflectivity (can be confirmed by Fig. 6). On further increase in the energy it increases and becomes zero at high energy limits.

The calculated refractive index and extinction coefficient of $Al_{1-x}Ga_xN$ at x = 0, 0.25, 0.50, 0.75 and 1.0 are presented in Fig.4. A broad spectrum for $n(\omega)$ over a wide energy and frequency range is noted for these compounds. The spectrum of $n(\omega)$ closely follow $\varepsilon_1(\omega)$ [27]. It is clear from the figure that the refractive index of the material increases with the increase in the Ga concentration in AlN. Fig.5 shows the plot between $n(0)$ and bandgap. The plot confirms the inverse relation of refractive index with the bandgap of $Al_{1-x}Ga_xN$. Two different features can be observed in the refractive index in Fig. 4. First; a maxima can be observed in the form of a bump in the spectrum at a particular energy and, second; the maxima shifts to lower energy region with the increase in Ga concentration. Pure AlN with x = 0 has the broadest spectrum of $n(\omega)$ with a maximum value around 2.0 at 8.0 eV. At intermediate energies a few bumps appear and then the curves vanish at higher energies. The reason for these vanishing curves at higher energy or frequency is due to the fact that, beyond certain energy, the material can no longer act as transparent material and it absorbs high energy photons. Perhaps the high bandgap of AlN makes it transparent up to high energy and frequency range [9]. However by increasing the concentration of Ga the appeared bumps in AlN changes to sharp peaks. The peaks are clear in pure GaN. At this point the refractive index reaches to its maximum value of more than 2.0 for the peak. Beyond this peak, the value of $n(\omega)$ dissipates rapidly. GaN absorbs most of the high and intermediate energy photons and is transparent in the low energy region due to the smaller bandgap of GaN than AlN [8,9].

It is also clear from Fig.4 that the refractive index falls below unity at certain frequencies. Refractive index lesser than unity *($v_g=c/n$)* shows that the group velocity of the incident radiation is greater than c. It means that the group velocity shifts to negative domain and the nature of the medium changes from linear to non-linear. In other words the material becomes superluminal for high energy photons [39, 40].

A relationship between $k(\omega)$ and energy is also shown in Fig.4. The response of $k(\omega)$ to the varying constituent's concentrations of the $Al_{1-x}Ga_xN$ is closely matching to the response of $\varepsilon_2(\omega)$. This result shows a good agreement with the available literature [27]. The peak value shifts to lower energies as x increases from 0 to 1.

Frequency dependent reflectivity, $R(\omega)$, versus energy for $Al_{1-x}Ga_xN$, is shown in Fig.6. Peaks in the figure show that for each concentration there is a maximum value of reflectivity. The maximum lie in the energy range 13 - 20 eV and arises from the inter band transition. While, the minimum of reflectivity occurs in the energy range 8 - 10 eV is due to the collective plasma resonance. The depth of the plasma resonance can be determined by the imaginary part of the dielectric function [41]. A more interesting aspect of the figure is the shift of the peak value towards lower energies with the increase in the Ga concentration. The material possesses reflectivity in wide energy and frequency range. For pure AlN the maximum value of reflectivity is 0.5, occurs at 17 eV and reduces to 0.46 at 13 eV for pure GaN. Hence, the reflectivity of $Al_{1-x}Ga_xN$ varies with the Ga concentration. This characteristic of the compound makes it suitable for Bragg's reflector in various wave lengths [42, 43]. Since the peak values of reflectivity in $Al_{1-x}Ga_xN$ changes with Ga concentrations, hence Bragg's reflector in the desired wavelength can be made and controlled by the Ga concentrations.

Frequency dependent absorption coefficient for $Al_{1-x}Ga_xN$ ($0 \leq x \leq 1$) is shown in Fig.7. The variation in frequency is expressed in terms of energy; ranging from 0 up to 30 eV. From the figure it is clear that the peak correspond to maximum absorption coefficient is shifted towards lower energies as the Ga concentration increases. Furthermore the maximum value drops from $300*10^4$ $cm^{-1}$ (for pure AlN) to $250*10^4$ $cm^{-1}$ (for pure GaN). From the figure, two interesting features of the absorption coefficient can be noted; one is the maximum value of the absorption coefficient shifts from 16 eV in pure AlN to 13 eV in pure GaN, while the second interesting feature is the appearance of the second peak with the increase in Ga concentration. The second peak is not observed in pure AlN, while two peaks are clear in GaN. These two peaks in the absorption coefficient explain the absorption of maximum light at two different wave lengths. Due to this property of the material, it can be used for wavelength filtering purposes in those regions.

## IV. CONCLUSION

Density functional calculations are used to study optoelectronic properties of $Al_{1-x}Ga_xN$ ($0 \leq x \leq 1$). Optical parameters like index of refraction, dielectric constant, reflectivity and absorption coefficient are strongly dependent on the concentration of Ga in the $Al_{1-x}Ga_xN$ crystal. Zero frequency limit of dielectric function, refractive index and reflectivity decreases with the increase of Ga in $Al_{1-x}Ga_xN$. The material becomes superluminal for photon energies larger than 14 eV, where the refractive index drops below 1. The peak value of the absorption coefficient and reflectivity shifts towards lower energy with the increase in Ga concentration, while two prominent peaks appear in the imaginary part of the dielectric function with Ga in AlN and the peak values enhances with the increase in Ga. The prominent variations in the optical parameters in the energy range 3 – 15 eV with Ga concentration in $Al_{1-x}Ga_xN$ makes it suitable for optical devices in the major parts of visible and UV spectrum.

**Acknowledgements**

Prof. Dr. Keith Prisbrey, MSE, University of Idaho and Prof. Dr. Nazma Ikram, Ex. Director, Center for Solid State Physics, Punjab University are highly acknowledged for their valuable suggestions.

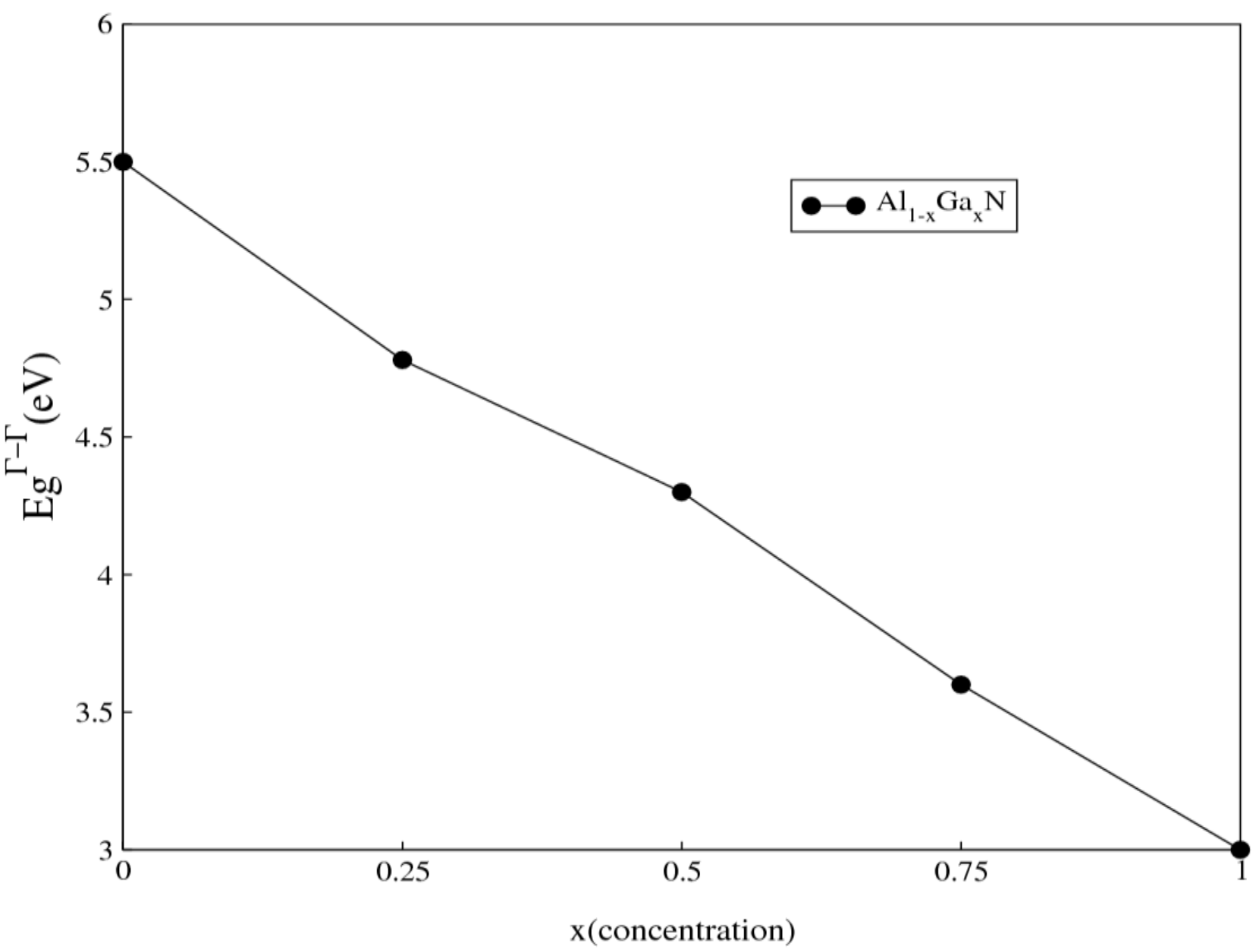

**Fig.1** Variation of direct bandgap with the concentration of x in $Al_{1-x}Ga_xN$ ($0 \le x \le 1$)

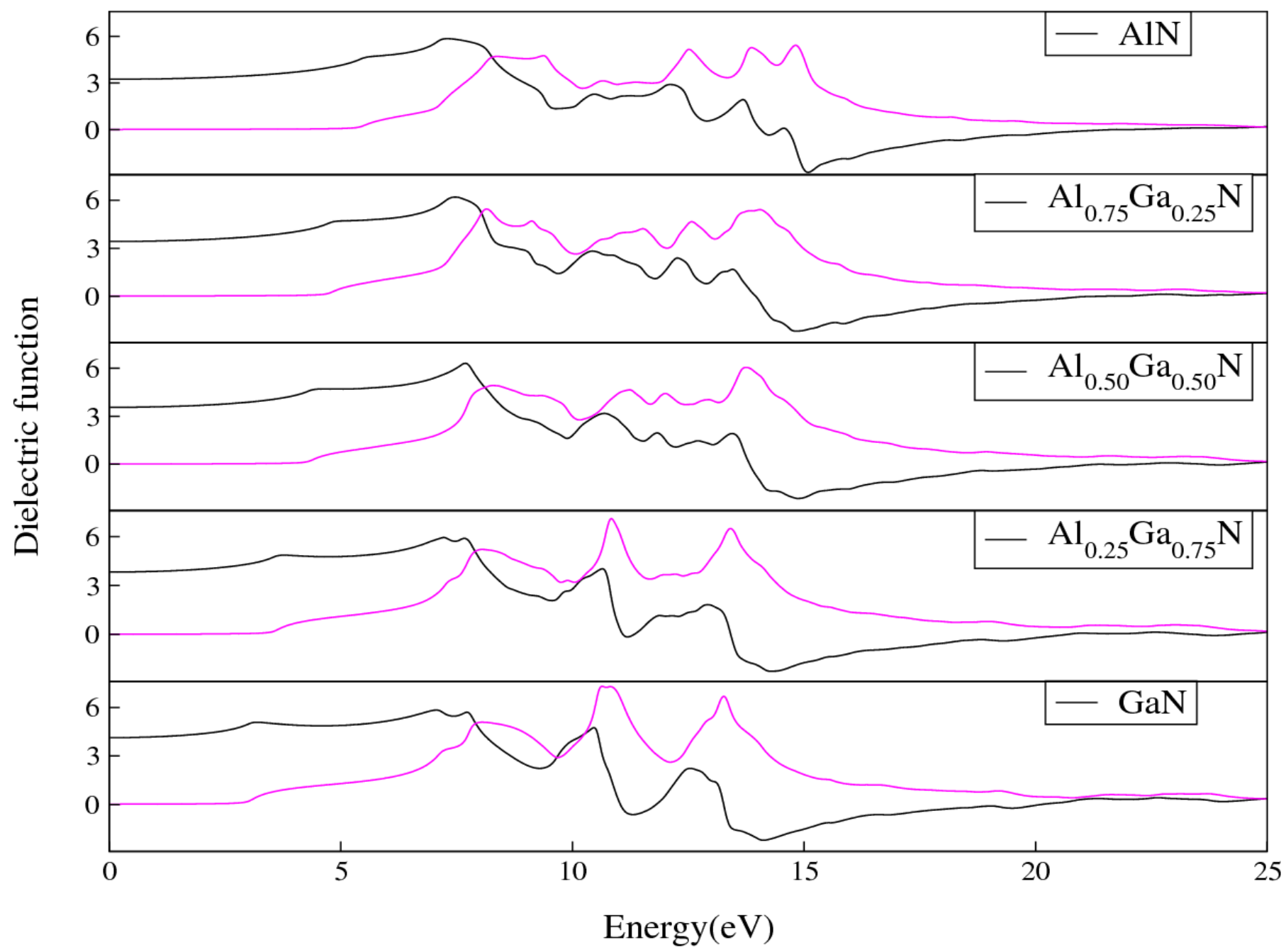

**Fig.2** Frequency dependent dielectric functions of $Al_{1-x}Ga_xN$ ($0 \leq x \leq 1$)

(Real part: black lines, imaginary part: magenta lines)

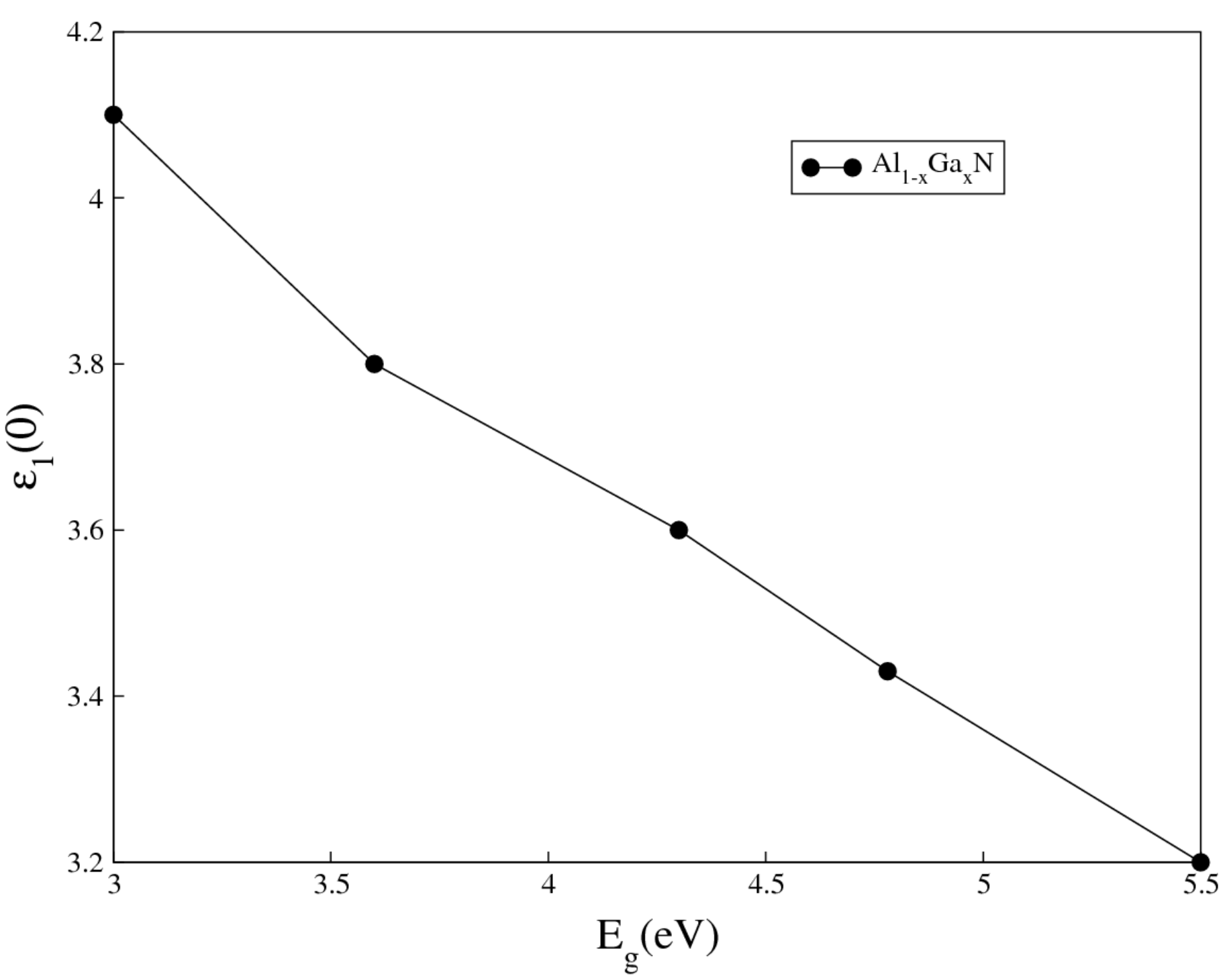

**Fig.3** Zero frequency limit of dielectric functions vs. direct bandgap of $Al_{1-x}Ga_xN$ ($0 \leq x \leq 1$)

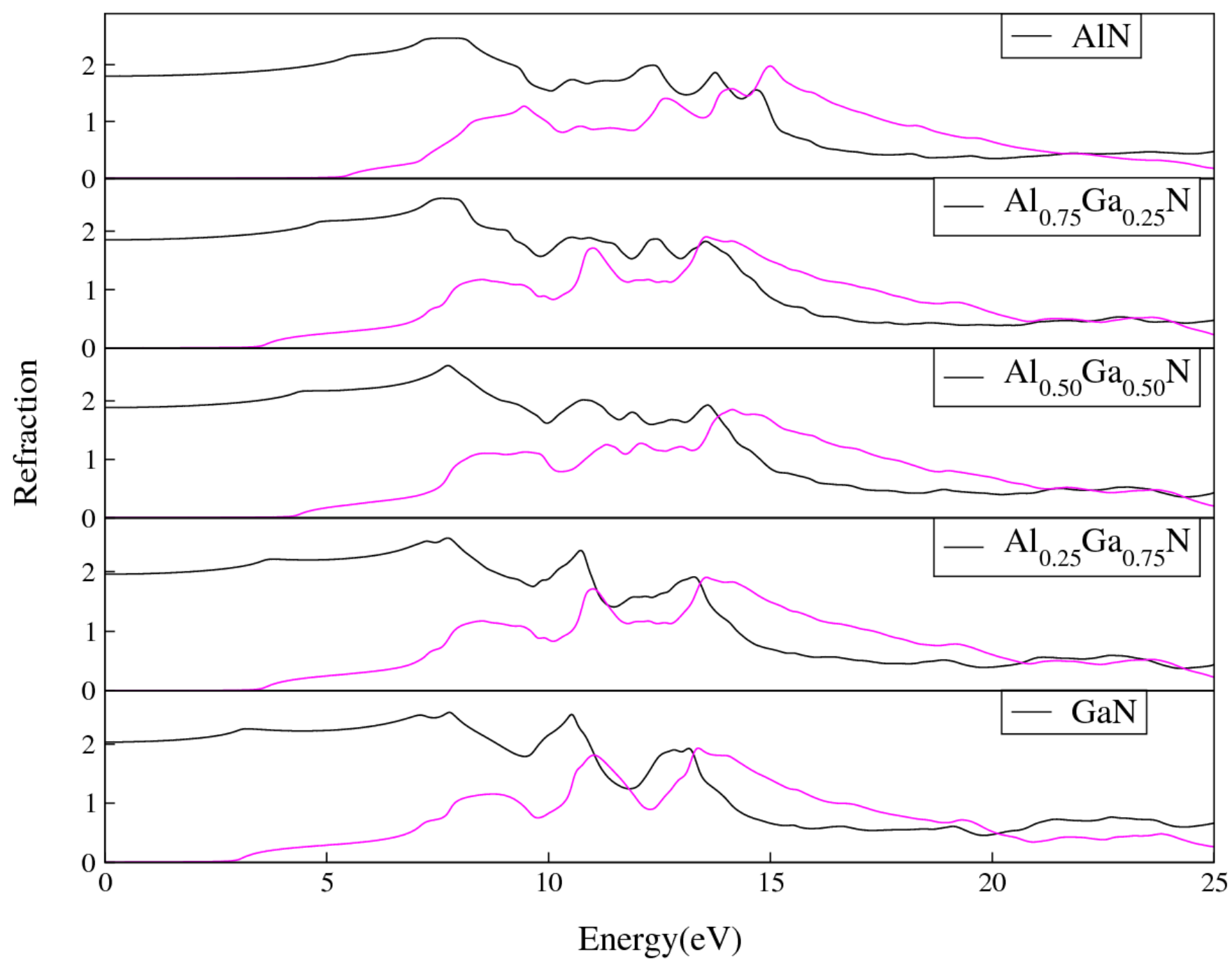

**Fig.4** Frequency dependent refractive indices of $Al_{1-x}Ga_xN$ (Real part: black lines, imaginary part: magenta lines)

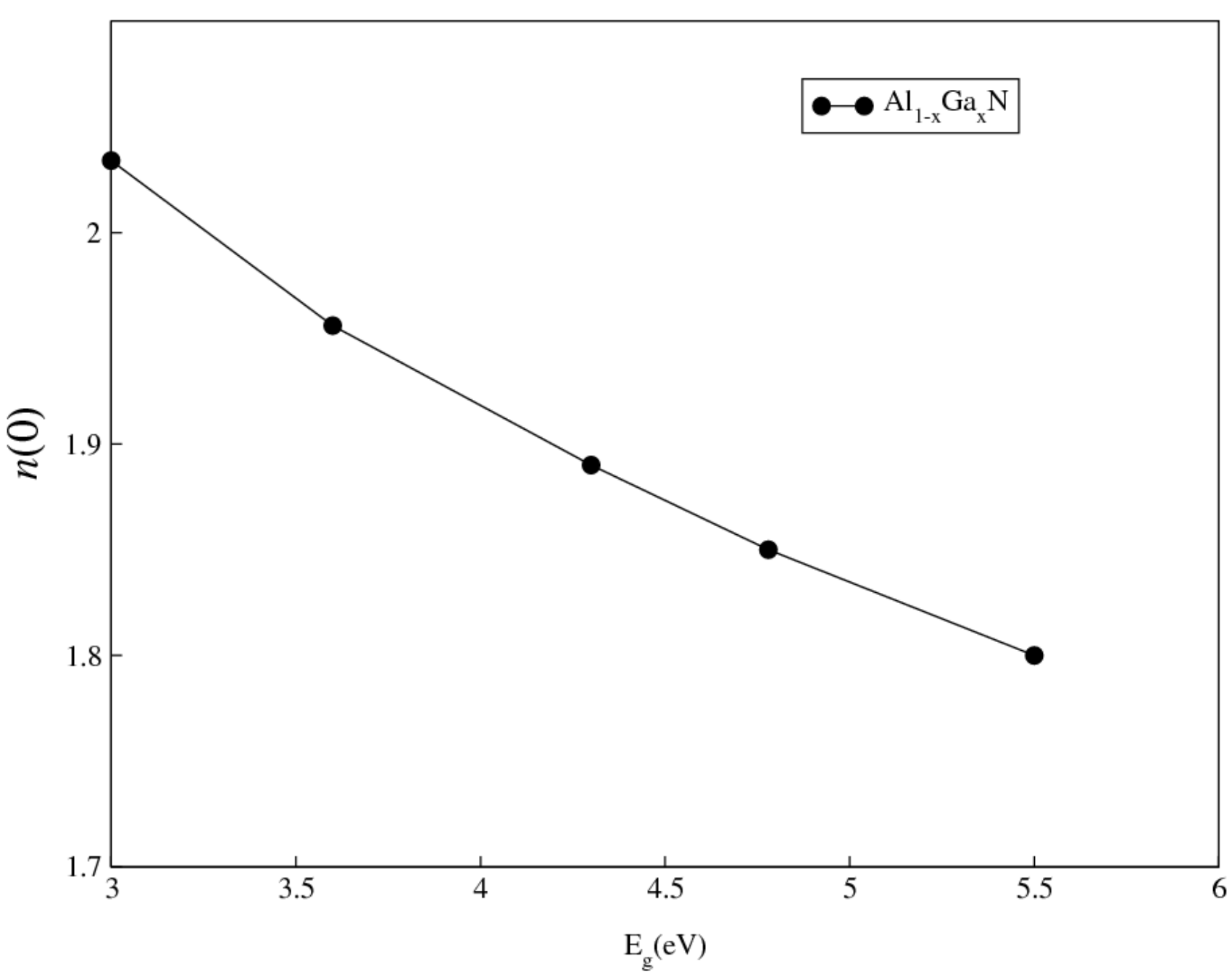

**Fig.5** Variation of static refractive index with bandgap of $Al_{1-x}Ga_xN$ ($0 \leq x \leq 1$)

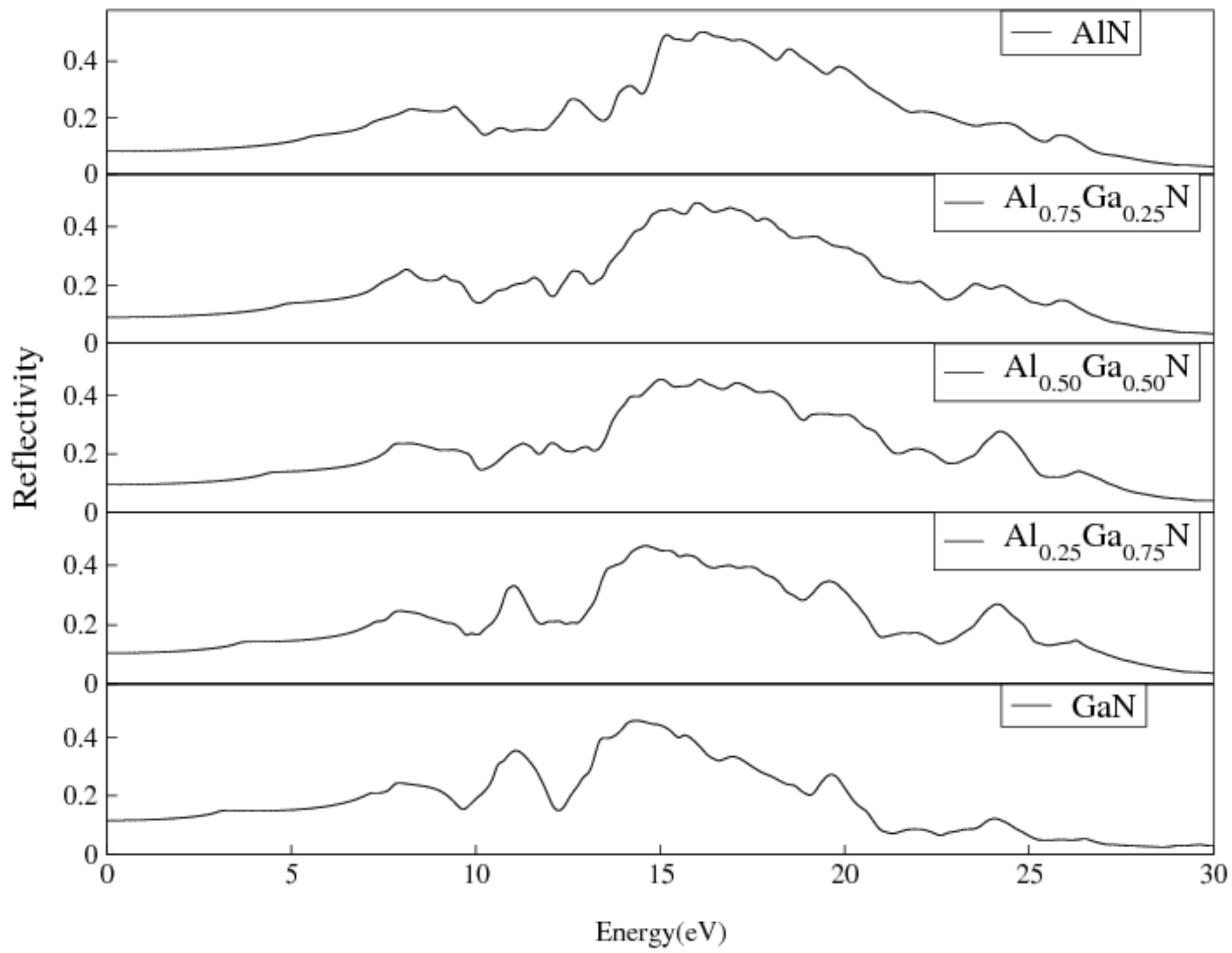

**Fig.6** Frequency dependent reflectivity of $Al_{1-x}Ga_xN$ ($0 \le x \le 1$) (Real part: black lines, Imaginary part: magenta lines)

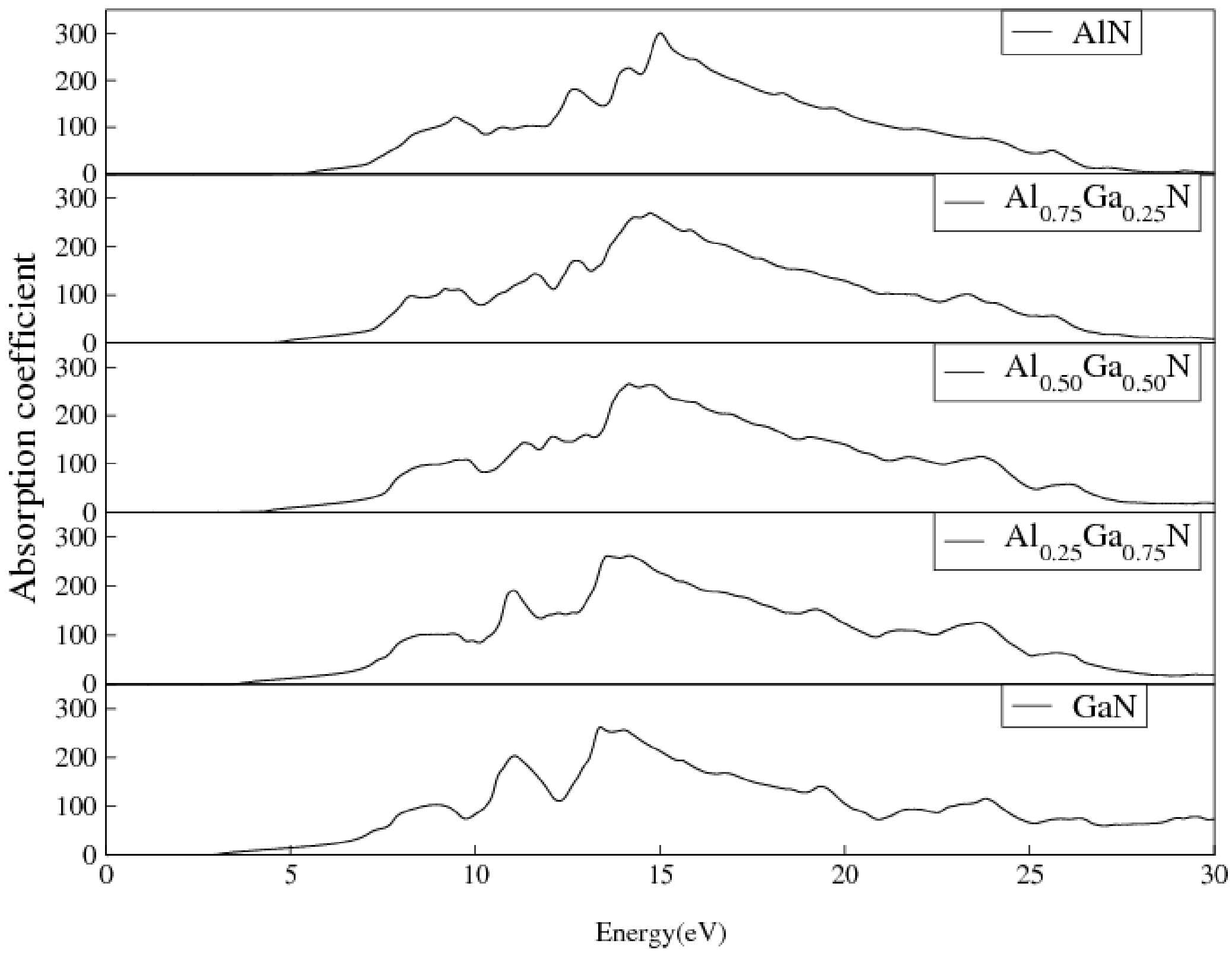

**Fig.7** Frequency dependent absorption coefficient of $Al_{1-x}Ga_xN$ ($0 \leq x \leq 1$)